\documentclass[12pt]{article}
\usepackage{epsf}                                                               
\newcommand{\R}{\rm I\kern-.2emR}
\newcommand{\C}{\rm \kern.25em\vrule height1.4ex
 depth-.12ex width.06em\kern-.31em C}
\newcommand{\N}{{\rm I\kern-.16em N}}
\newcommand{\Z}{{\rm Z\kern-.35em Z}}
\newcommand{\bee}{\begin{equation}}
\newcommand{\ee}{\end{equation}}
\newcommand{\ba}{\begin{array}}
\newcommand{\ea}{\end{array}}
\newcommand{\bea}{\begin{eqnarray}}
\newcommand{\eea}{\end{eqnarray}}

\begin{document}
\begin{flushright}                                                              
MPI-PhT 96-86  \\
\end{flushright}                                                                
\bigskip\bigskip\begin{center}
\renewcommand{\thefootnote}{\fnsymbol{footnote}}
{\bf The Problem of Asymptotic Freedom}\footnote[1]
{Paper presented by E.Seiler at the 28th International Conference on 
High Energy Physics}
\end{center}  
\vskip 1.0truecm
\centerline{Adrian Patrascioiu}
\centerline{\it Physics Department, University of Arizona}
\centerline{\it Tucson, AZ 85721, U.S.A.}
\vskip5mm
\centerline{and}
\vskip5mm
\centerline{Erhard Seiler}
\centerline{\it Max-Planck-Institut f\"{u}r Physik}
\centerline{\it  -- Werner-Heisenberg-Institut -- }
\centerline{\it F\"ohringer Ring 6, 80805 Munich, Germany}
\bigskip \nopagebreak \begin{abstract}
\noindent
There is a growing body of evidence that the running of
$\alpha_s$ predicted by perturbation (PT) theory is not correctly 
describing the accelerator experiments at the highest energies. A natural explanation
is provided by the authors' 1992 proposal that in fact the true running
predicted by the nonperturbatively defined lattice QCD is different, leading to
an ultraviolet fixed point near $\alpha_s=.1$.
It is explained how this can be understood from the fact that the 
conventional perturbative method is ambiguous and does not provide the
correct asymptotic expansion. It is pointed out that there is a large
amount of lattice data that are supporting this scenario rather than the
conventional one.
\end{abstract}
\vskip 5mm
\vskip4mm \noindent
{\bf 1.Introduction}
\vskip2mm
Asymptotic Freedom (AF) is reputed to be:

\noindent
(1) a property of nature (the strong force)

\noindent
(2) a property of the theory called Quantum Chromodynamics (QCD),
expressed by the familiar formula for the running coupling constant
\bee
\alpha_s(\mu)={12\pi\over (33-2n_f)\ln{\mu^2\over\Lambda^2}}
\Biggl[1-{6(153-19n_f)\over (33-2n_f)^2}
{\ln\ln{\mu^2\over\Lambda^2}\over\ln{\mu^2\over\Lambda^2}}\Biggr]
+ ...
\ee
Both properties are a priori logically independent and require 
experimental and theoretical confirmation, respectively. This would also
help answer the important question whether QCD indeed describes
correctly the strong interaction.

For most particle physicists these appear to be questions that were
settled long ago; we want to emphasize that both of these two properties
mentioned above are far from being established beyond any reasonable
doubt and we think there are good reasons to expect that AF
in the strict sense neither holds in QCD nor in nature. 

We will point out that Lattice QCD (LQCD), if one approaches its results 
with an unprejudiced mind, in fact suggests the existence of a 
nonzero fixed point for the strong coupling constant $\alpha_s$, in 
conflict with the property of AF derived from perturbative QCD (PQCD)
more than 20 years ago \cite{GW,P}. Assuming that QCD indeed describes
the strong interaction correctly,
this in turn predicts that in nature $\alpha_s$ will run to a fixed
point $\alpha_{fp}>0$, and that at high energies the running of
$\alpha_s$ is slower than the PQCD prediction.

\vskip4mm \noindent
{\bf 2.The Need for a Nonperturbative Definition of QCD}
\vskip2mm
QCD is a field theory of quark and gluon fields; but nature does not 
know any particles identifiable with quarks and gluons, only hadrons,
which are considered as permanently bound states of the former. To
substantiate this claim, i.e. to show that quarks and gluons are
permanently bound into states corresponding to the known hadrons, one 
obviously needs a nonperturbative definition of the theory. To this day 
the only serious and useful such definition is provided by LQCD 
\cite{Wilson}. In view of the undoubted successes of LQCD in
describing the low energy phenomenology of hadrons we accept as a working
hypothesis that QCD, defined nonperturbatively as the continuum limit of 
lattice QCD, in fact correctly describes the strong force.

AF, on the other hand, is a property derived exclusively in the
framework of {\it perturbative} QCD (PQCD) \cite {GW,P}.
For many phenomenologists, QCD has become synonymous with PQCD. It has to
be stressed that this standpoint is not satisfactory for two reasons: \\
1.~PT is an algorithm that produces for a physical quantity, such as
\bee
R(Q)={\sigma(e^+e^-\to\mbox{hadrons})\over(\sigma(e^+e^-\to\mu^+\mu^-)}
\ee
a formal power series in $\alpha_s(Q)$:
\bee
R(Q)\sim\sum_{k=0}^\infty c_k\alpha_s(Q)^k;
\ee
the radius of convergence of this formal series is generally expected to 
be zero. The meaning of eq.(2) is therefore a priori unclear.
PQCD does not really predict definite numbers, but sequences of
numbers, depending on where one decides to truncate the divergent series 
(2). Generally one stops at a low order because the computation of high 
orders is beyond human capablities anyway, and hopes that the truncated 
series represents the `truth' to a good approximation. That there is a 
problem becomes clear if one assumes that a fairy would give as 
{\it all} the PT coefficients. We could not extract in an unambiguous way
a definite number as the `sum' of this series!

Many people would object and say that we should try to use some 
prescription like Borel or Borel-Pad\'e summation. But even if these
procedures would in fact produce a number as the `sum' of the series,
in the absence of a nonperturbative definition this would be just an
ad hoc procedure and one could not even meaningfully ask if the answer
is `right'.

2. Of course it is generally said that PT is an {\it asymptotic series}.
To give meaning to this statement, one has to accept a nonperturbative 
definition of QCD,
such as the continuum limit of LQCD. Then one can ask whether
PT is indeed asymptotic to this nonperturbatively defined theory,
i.e. (for the example of $R(Q)$):
Are there numbers (or maybe functions of $Q$) $d_k$, $k=1,2,3,...$
such that the following inequality holds for all $N$
\bee
|R(Q)-\sum_{k=0}^N\alpha_s(Q)^k c_k|<d_{N+1}\alpha(Q)^{N+1}  ?
\ee

\vskip4mm \noindent
{\bf 3.Can Perturbation Theory be Trusted?}
\vskip2mm
LQCD is defined by the (infinite volume limit) of the Gibbs measure
\bee
d\mu_\Lambda={1\over Z}\exp(-\beta S_\Lambda)
\ee
with $\beta=6/g_o^2$.

Accepting now LQCD as our nonperturbative definition, we ask about
the trustworthiness of standard PT. On the lattice PT is really a saddle
point expansion around an ordered state (in gauge theories a state with
vanishing gauge field strength). While in a fixed lattice volume, by 
making the (bare) coupling small, one can make the deviations from
such an ordered state small, provided one fixes the gauge completely,
in an infinite system this is not the case. There are old arguments
\cite{Pat} and rigorous results \cite{SY} that show that in the
(complete) axial gauge there are still arbitrary large fluctuations
no matter how small the coupling. PT --  which always has to be done in 
an arbitrarily large lattice volume if we want to approach the continuum,
even if we are in a finite physical volume --  is thus an expansion in
a quantity that is not small. (In the Landau gauge, the problem appears
in a different guise: the integration is cut off arbitrarily close
to the saddle point due to the presence of the Gribov horizon
\cite{20er}).
So why could one expect it to give the right answer, even if in certain 
quantities the volume divergences seem to cancel term by term?

The presence of these large fluctuations can be understood from the
existence of excitations of arbitrarily low energy (action) that
lead the system arbitrarily far from any assumed ordered state at
arbitrarily low cost in energy. These low-lying excitations were dubbed
`superinstantons' by us \cite{sigauge}
in order to emphasize that at weak coupling they
will be present much more frequently than instantons, which have a
finite excitation energy over the classical vacuum. The true QCD vacuum
really has to be imagined as a {\it condensate of superinstantons}.

That something is indeed dubious with the usual procedure of performing
a saddle point expansion with an infrared (IR) cutoff (such as a
finite volume) and then removing the cutoff term by term, can be seen
explicitly if we study the dependence of PT on boundary conditions 
(b.c.). It is a general fact that precisely due to the large fluctuations
that put PT into question, the true, nonperturbatively defined
infinite volume limit becomes insensitive to b.c.. 

On the other hand, we have shown in \cite{sigauge} that the PT 
coefficients
in fact do have different infinite volume limits for different b.c.! 
This affects even the Callan-Symanzik function $\beta_{CS}$. So
PT gives ambiguous results, and certainly has to fail for some b.c. Nobody
knows for which b.c. it might be right. Most likely it does not give
the right answer for any simple b.c.; to get the right answer, in
addition to the trivial saddle point, all the low lying
superinstantons (almost degenerate with the trivial saddle point) have to
be taken into account to get the true asymptotic expansion. How this
is to be done concretely, remains an unsolved problem.

\vskip4mm \noindent
{\bf 4.Numerical and Analytic Evidence}
\vskip2mm
For LQCD the perturbatively computed Callan-Symanzik function 
$\beta_{CS}$ has the form
\bee
\beta_{CS}=b_og_o^3+b_1g_o^5+ ...
\ee
It leads to the prediction that any dynamically generated mass will be
proportional asymptotically for $\beta\to\infty$ to
\bee
\Lambda_L(\beta)=\exp(-{\beta\over 12b_o})\beta^{b_1/2b_o^2}
\ee
(asymptotic scaling). Numerical studies have found consistently that
this is violated as one enters the scaling region, where the correlation
length begins to grow and continuum behavior should begin to be seen.
This is the so-called dip in $\Delta\beta$, the change in $\beta$
correpsonding to a doubling of the correlation length. This `dip'
occurs likewise in the $2D$ toy analogues
of QCD such as the $O(N)$ nonlinear $\sigma$ models. The deviation is
always such that the correlation length increases faster with $\beta$
that eq.(5) predicts, just as if the theory was really approaching
a critical point. 

There have been some proposals to argue that phenomenon away. First
of all people have tried to compute masses for $\beta$ values so large
that it is impossible to work on `thermodynamic' lattices (much larger
than the correlation length). In essence these proposals are using small
lattices and try to extrapolate what they find there to thermodynamic 
ones (this is the common feature of the so-called Monte Carlo
Renormalization Group (MCRG) and Finite Size Scaling (FSS)). What has been
found using these ideas is that apparently the `dip' disappears at larger
$\beta$ values and the PT $\beta$ function really seems to describe what
is happening. But this is merely a reflection of the fact that the
small lattices are actually very much ordered and don't admit the large
fluctuations that are necessary to describe the continuum behavior
(see \cite{Kimcom} for a detailed discussion of an example).
It has also been noted long ago by Gutbrod \cite{Gutbrod}
that the `dip' deepens, as data from larger lattice are introduced.
In our opinion there is not a dip, but a zero of $\Delta\beta$ at
some $\beta$.

Another popular fix is to replace $\beta$ in eq.(4) by an `effective'
quantity that asymptotically becomes equal to $\beta$ ( such as in
the so-called energy scheme). This procedure appears to be completely 
ad hoc; it is not surprising that by a suitable transformation 
$\beta\to\beta_{eff}$ one can achieve perfect asymptotic scaling over 
any finite range.

A typical example of the `dip' is provided by Fig.\ref{Bielefig}
produced with data of the Bielefeld group \cite{Biele},
actually not referring to a mass, but the deconfinement temperature:
                                                                                
\begin{figure}[htb]                                                             
\centerline{\epsfxsize=11cm\epsfbox{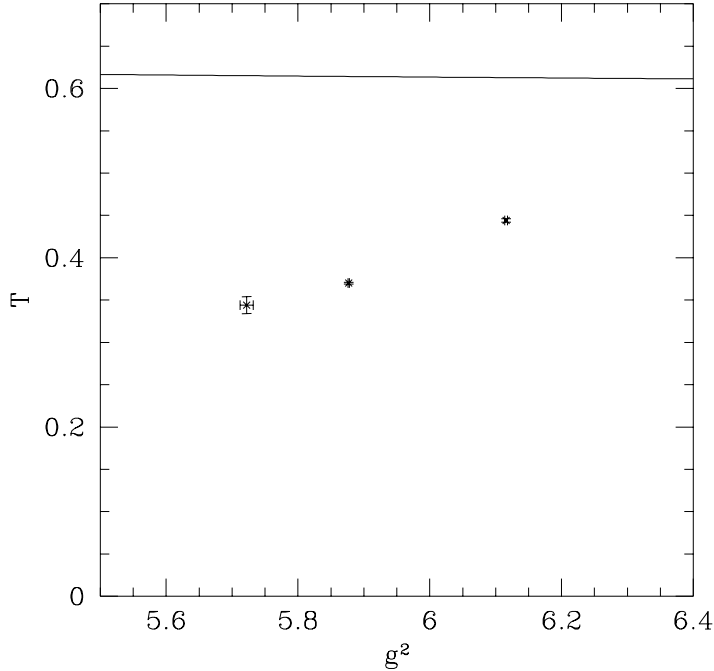}}
\caption{The `dip' as obtained from the deconfinement temperature;
the solid line is the 2 loop PT prediction and the data are taken from 
10].}
\label{Bielefig}
\end{figure}                                                                    

It should be noted that the three points are obtained by
some extrapolation from a lattice that is too small, so it is not certain
what the true infinite volume answer is; we expect the points eventually
to move down as computing power and lattice sizes increase. In fact 
earlier studies, using smaller lattices \cite{Gottlieb, Christ},
claimed that asymptotic scaling sets in around $\beta=6.0$, which is 
obviously not borne out by the Bielefeld results.

Bali et al.\cite{Bali} gave another example; they find (for the
$SU(2)$ pure gauge model at $\beta=2.5$) a value for $\beta_{CS}$ that 
is only $63\%$ of the PT prediction.

Now to the analytic evidence: as mentioned, LQCD has a little brother 
in the $2D$ $O(N)$ models for $N>2$. These models also share the property
of perturbative asymptotic freedom and large fluctuations that make PT 
both suspect and ambiguous. For those models we developed an argument 
based on percolation theory
leading to the conclusion that there is a critical point (fixed point)
at nonzero coupling, in drastic conflict with the predictions of PT.
This chain of arguments, while not completely rigorous, has remained
without serious challenge over the more than four years since it was 
presented publicly \cite{ANP,Lat92}.
(For LQCD this kind of argument cannot yet be made, just like there is
not yet a good cluster algorithm for gauge models; but we think, like
the rest of the community, that the two types of models are really
behaving the same way).

So we think there is good reason to doubt the conventional wisdom and
accept the existence of a phase transition of LQCD at nonzero coupling.

\vskip2mm
\vskip4mm \noindent
{\bf 5. Consequences}
\vskip2mm

Finally let me say something about the consequences of the proposed
fixed point in the running of $\alpha_s$. These were first pointed out
by us in \cite{expdev}. It is quite clear that such a fixed point means
that eventually the running of $\alpha_s$ has to be slower than what
PT predicts; this may also be expressed by saying that
$\Lambda_{QCD}$ is not aconstant but an increasing function of the
momentum $Q$.

At which energy scale are deviations expected to be seen? Is it
something like the Planck or GUT scale, or is an accessible energy?
The answer to this questions in some sense decides if we are talking about
physics or metaphysics.

To find an answer, one first has to estimate the critical lattice 
coupling. From an analysis of the numerical
results it seems plausible to assume $\beta_{crit}\approx 7-10$.
This has to be converted into a mass scale (i.e. the size of the UV 
cutoff), and for lack of anything better we use PT for the moment to 
estimate it. This leads, maybe surprisingly, to a not so huge scale
of 1 to 3 TeV. Since of course the existence of the assumed critical 
point means that PT really has to fail drastically at this value of
$\beta$, we expect drastic deviations from perturbative running at
the scale of 1 to 3 TeV. But of course these deviations will set in 
gradually, and therefore we predicted that deviations will be seen at
1 TeV {\it or less}.

A naive ansatz for the variation of $\Lambda$ with
$Q$ is to replace $\Lambda^2$ by $\Lambda^2+Q^2$. This ansatz can be 
shown to correspond to mean field critical behavior; mean field behavior
is expected to hold in $4D$ LQCD at least in one instance, namely in
the PCAC relation. With this simple minded ansatz, the true running of 
$\alpha_s$ might look as in Fig.\ref{alphafig}:
                                                                                
\begin{figure}[htb]                                                             
\centerline{\epsfxsize=11cm\epsfbox{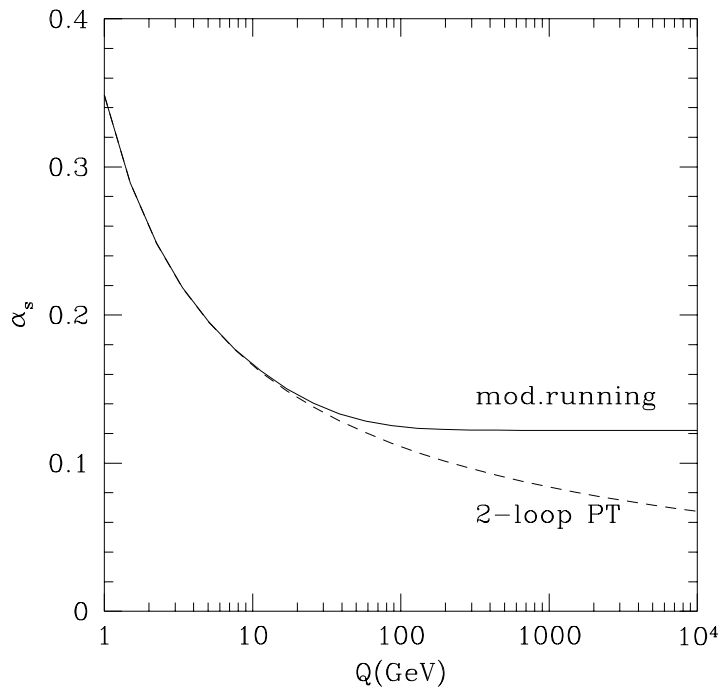}}
\caption{Possible modified running of $\alpha_s$}
\label{alphafig}
\end{figure}                                                                    

Soon after we made our prediction, the LEP results seemed indeed to
indicate such a deviation: instead of $\alpha_s(m_Z)=.113$ as predicted
by Altarelli \cite{Altarelli} on the basis of lower energy results, the 
central LEP value is now .123 . Of course there is no consensus on the
significance of the deviation, which is (depending on whom you ask) at the
2 or  $3\sigma$ level. But this has not deterred people to interprete it
as a sign of the existence of light gluinos \cite{Hans,Ellis} or more
generally as signifying the presence of {\it new physics} \cite{Misha}.
The obvious and much more economical explanation that the actual running 
is different from what PT says, seems to be impalatable to people.

(It should be noted, however, that at least one team has picked up on the
idea of a fixed point in $\alpha_s$ \cite{SS}. The authors checked whether
the deep inelastic scattering (DIS) data are compatible with a fixed
point and found that they are. Since we know of no reaction of the
community to this analysis, we don't know if there are any valid
objections to it).

A similar (social) phenomenon could be observed again this year with
the publicity surrounding the CDF results on enhanced jet production
at very large $p_T$. Dramatic claims of {\it new physics} were made, but
the obvious possibility of a slowly running $\alpha_s$ remained 
unexplored.

All these questions are too important to remain uninvestigated.
Experimental physicists should be convinced that it is worthwhile to
provide better results on the running of $\alpha_s$ (supposedly CLEO
could provide a very good value at a `low' energy). Lattice theorists
should at least recognize that there is an important and challenging
issue to be resolved, instead of simply taking for granted what most
people believe. The data provide strong hints that something is going on,
and it is not helpful to argue these hints away by changing the rules of
the game, introducing ad hoc fixes like the `energy scheme' or to produce
a host of Monte Carlo data at small lattices that could just as well be
be produced without much effort by using lattice PT, and which therefore
do not make as any smarter as to the nonperturbative behavior of LQCD.


\begin{thebibliography}{99}
%
\bibitem{GW} D.Gross and F.Wilczek, {\sl Phys.Rev.Lett} {\bf 30} (1973)
1181.
%
\bibitem{P} H.D.Politzer, {\sl Phys.Rev.Lett.} {\bf 30} (1973) 1346.
%
\bibitem{Wilson} K.Wilson, {\sl Phys.Rev.} {\bf D10}, (1974) 2445.
%
\bibitem{Pat} A.Patrascioiu, {\sl Phys.Rev.Lett.} {\bf 54} (1985) 2292.
%
\bibitem{SY} B.Simon and L.G.Yaffe, {\sl Phys.Lett.}
{\bf 115B} (1982) 145;
M. L\"uscher, {\it Absence of Spontaneous Gauge
Symmetry Breaking in Hamiltonian Lattice Gauge Theories}, preprint DESY-77/16.
%
\bibitem{20er} D.Zwanziger, {\sl Nucl.Phys.} {\bf B378} (1992) 525.
%
\bibitem{sigauge} A.Patrascioiu and E.Seiler,
{\sl Phys.Rev.Lett.} {\bf 74} (1995) 1924.
%
\bibitem{Kimcom} A.Patrascioiu and E.Seiler,
{\sl Phys.Rev.Lett.} {\bf 73} (1994) 3325; {\sl Phys.Rev.Lett.} {\bf 76}
(1996) 1178.
%
\bibitem{Gutbrod} F.Gutbrod, {\sl Phys.Lett.} {\bf B 186} (1987) 389;
{\sl Z.Phys.} {\bf C 37} (1987) 143; {\sl Z.Phys.} {\bf C 55} (1992) 463.
%
\bibitem{Biele} G.Boyd, J.Engels, F.Karsch, E.Laermann, C.Legeland,
M.L\"utgemeier and B.Petersson,
{\it Thermodynamics of $SU(3)$ Lattice Gauge Theory},
hep-lat/960207.
%
\bibitem{Gottlieb} S.A.Gottlieb, J.Kuti, D.Toussaint, A.D.Kennedy,
S.Meyer, B.J.Pendleton and R.L.Sugar, {\sl Phys.Rev.Lett.} {\bf 55}
(1985) 1958.
%
\bibitem{Christ} N.Christ and A.E.Terrano, {\sl Phys.Rev.Lett.} {\bf 56}
(1986) 111.
%
\bibitem{Bali} G.Bali, C.Schlichter and K.Schilling, {\sl Phys.Lett.}
{\bf B 363} (1995) 196.
%
\bibitem{ANP} A.Patrascioiu, {\it Existence of Algebraic Decay in
non-Abelian Ferromagnets}, University of Arizona preprint AZPH-TH/91-49.
%
\bibitem{Lat92} A.Patrascioiu and E.Seiler, {\it Percolation Theory and
the Existence of a Soft Phase in 2D Spin Models}, {\sl Nucl.Phys.B.(Proc.
Suppl.)} {\bf 30} (1993) 184.
%
\bibitem{expdev} A.Patrascioiu and E.Seiler,
{\it Expected Deviations From Perturbative QCD at 1 TeV or Less},
preprint MPI-Ph/92-18 and AZPH-TH-92-06;
{\it Observable Deviations from Perturbative QCD},
in: {\it Rencontre de Physique de la Vall\'ee D'Aoste}, M.Greco(ed.), 
Fronti`eres 1992.
%
\bibitem{Altarelli} G.Altarelli, {\sl Ann.Rev.Nucl.Part.Sci.} {\bf 39}
(1989) 357.
%
\bibitem{Hans} M.Jezabek and J.H.K\"uhn, {\sl Phys.Lett.}
{\bf B 301} (1993) 121.
%
\bibitem {Ellis} J.Ellis, D.V.Nanopoulos, Douglas A.Ross, {\sl Phys.Lett.}
{\bf B305} (1993) 375.
%
\bibitem{Misha} M.Shifman, {\it The Case of $\alpha_s$: $Z$ versus Low
Energies or How Nature Prompts us of New Physics},
TPI-MINN-95/32-T, UMN-TH-1416-95, hep-ph/9511469.
%
\bibitem{SS} A.V.Sidorov and D.B.Stamenov, {\sl Phys.Lett.}
{\bf B 357} (1996) 423.
%
\vfill\eject

\end{thebibliography}
\end{document}